\documentclass{PoS}
\usepackage{bm}
\usepackage{amsmath}

\title{Cosmological Helical Hypermagnetic Fields and Baryogenesis}

\ShortTitle{Cosmological Helical Hypermagnetic Fields and Baryogenesis}

\author{\speaker{Kohei Kamada}\thanks{Present affiliation: Research Center for the Early Universe (RESCEU),
Graduate School of Science, The University of Tokyo, Tokyo 113-0033, Japan}\\
        Center for Theoretical Physics of the Universe,
Institute for Basic Science (IBS), Daejeon, 34126, Korea\\
        E-mail: \email{kkamada@ibs.re.kr}}

\abstract{
Recent gamma-ray observations of TeV blazars exhibits the deficits of the secondary GeV cascade photons. 
This suggests the existence of the intergalactic magnetic fields, which may have a primordial origin. 
One of the mechanisms that can produce primordial magnetic fields is so-called 
the chiral plasma instability, where the (hyper) magnetic fields are destabilized 
when a large chiral asymmetry exists in
the high-temperature plasma in the early Universe. 
We argue that such a large
chiral asymmetry can be produced through the GUT baryogenesis. 
Note that the chiral asymmetry is a good conserved quantity at high temperature when the
Yukawa interaction is weak enough. 
We also point out that the generated hypermagnetic fields are maximally helical, 
and hence baryon and lepton asymmetry is inevitably produced through the chiral anomaly in the 
Standard Model through U(1)$_Y$ gauge interaction at the electroweak symmetry breaking.
Consequently, the magnetic fields suggested by the blazar observations over-produce baryon asymmetry. 
Thus the chiral plasma instability alone cannot be responsible for the intergalactic magnetic fields 
but can be responsible for the baryon asymmetry of the Universe. 
 In other words, GUT
baryogenesis without $B$-$L$ asymmetry generation is revived as a
viable baryogenesis scenario, which otherwise has been thought
to suffer from $B$+$L$ washout by sphalerons. This presentation is based on the work~\cite{Kamada:2018tcs}. }

\FullConference{The 39th International Conference on High Energy Physics (ICHEP2018)\\
		4-11 July, 2018\\
		Seoul, Korea}

\begin{document}

\section{Cosmological magnetic fields}

Primordial magnetic fields have been of interest for a long time as the origin of the seed magnetic fields for the observed 
galaxy and galaxy cluster magnetic fields. 
Moreover, recent observations of gamma rays from TeV blazars identified the deficits of GeV cascade photons, 
which should accompany with the TeV photons, suggest the existence of the 
intergalactic magnetic fields. 
According to the latest constraint by Fermi LAT~\cite{Biteau:2018tmv}, the lower bound of the magnetic field strength today is roughly 
given by 
\begin{equation}
B_0 \gtrsim \left\{\begin{array}{ll}
10^{-16} {\rm G}, & \quad {\rm for} \quad \lambda_0 > 10^{-2} {\rm Mpc}, \\
10^{-16} {\rm G}\times  (\lambda_0/10^{-2} {\rm Mpc})^{-1/2}, & \quad {\rm for} \quad \lambda_0 < 10^{-2} {\rm Mpc},
\end{array}\right.
\end{equation}
where $\lambda_0$ is the present correlation length of magnetic fields. 
These observations motivate us to explore the possibilities of mechanisms to generate magnetic fields (magnetogenesis) 
in the early Universe, especially primordial ones such as inflationary magnetogenesis or that from cosmological phase transitions. 
However, at present there are no satisfactory models that can explain the lower bound of the intergalactic magnetic fields, 
and hence it is worth exploring other ideas to generate them and is interesting to investigate 
the phenomena induced by the magnetic fields in the early Universe regardless of 
their possibility to be responsible for the magnetic fields today.

\section{Chiral plasma instability}

Another interesting mechanism to produce magnetic fields is the chiral plasma instability~\cite{Joyce:1997uy}, 
which is very recently investigated in the numerical magnetohydrodynaimics (MHD) studies, {\it e.g.} Ref.~\cite{Schober:2017cdw}. 
This mechanism takes advantages of the chiral magnetic effect~\cite{Vilenkin:1980fu}, 
in which a electric current parallel to the magnetic field ${\bm B}$ and proportional to the chiral chemical potential $\mu_5$
is induced in the thermal plasma with nonzero chiral chemical potential, ${\bm J}_{\rm CME} =  (2\alpha/\pi) \mu_5 {\bm B}$, 
where $\alpha$ is the fine structure constant. 
In this circumstances, the Maxwell's equations in the Standard Model (SM) thermal plasma are modified as 
\begin{equation}
\frac{d{\bm B}_Y}{d \tau} = -{\bm \nabla} \times {\bm E}_Y, \quad {\bm \nabla} \times {\bm B}_Y = {\bm J}_Y,  \quad {\rm with} \quad 
{\bm J}_Y = \sigma_Y ({\bm E}_Y+{\bm v}\times {\bm B}_Y) + \frac{2 \alpha_Y}{\pi} \mu_{5,Y} {\bm B}_Y,  \label{max}
\end{equation}
where the subscript $Y$ represents that the variables are for the SM hypergauge interactions SU(1)$_Y$. 
$\sigma$ is the electric conductivity for the Ohm's current, ${\bm v}$ is the plasma velocity field and $\tau$ is 
the conformal time. 
For a constant $\mu_{5,Y}$ and a negligibly small ${\bm v}$, the mode equations for the circular polarization (helicity) of the hyper 
magnetic fields are given by 
\begin{equation}
\frac{d B_k^{Y\pm}}{d \tau} = -\frac{k}{\sigma_Y}\left(k \mp \frac{2 \alpha_Y}{\pi} \mu_{5,Y}\right) B_k^{Y\pm}. \label{mode}
\end{equation}
This clearly suggests that one of the helicity modes of the hyper magnetic fields feels instability for 
$k<2 (\alpha_Y/\pi) \mu_{5, Y}$ and is most efficient at $k_c =  (\alpha_Y/\pi) \mu_{5, Y}$.
As a consequence, the hyper magnetic fields are exponentially amplified with being maximally helical.  

Through the MHD ${\bm v}$ fields are also induced, and then the evolution equation~\eqref{mode} will be no longer hold. 
However, recent numerical MHD studies~\cite{Schober:2017cdw} showed that even when the ${\bm v}$ fields are induced 
the exponential growth of magnetic fields still continues (with slight suppression) until the 
magnetic field amplitude is saturated and then it is argued that the magnetic fields  will obey the inverse cascade process,
where the magnetic field strength and coherence length evolves as 
${\bm B}_Y \propto \tau^{-1/3}$ and $\lambda \propto \tau^{2/3}$. 
The saturation is determined by the conservation law between the chirality and magnetic helicity, 
coming from the chiral anomaly equation, $\mu_{5,Y} + 3 c_1 \alpha_Y h / (\pi T_i^2) = {\rm const}.$ with the helicity being 
$h \simeq \lambda |{\bm B}_Y|^2/2\pi$, $T_i$ being the initial temperature
and $c_1$ being the initial chiral asymmetry to right-handed electron asymmetry ratio,  
which also depends on the particle contents. 
That is, the final helicity just after the saturation is 
$h \simeq \pi T_i^2 \mu_{5,Y}^i / 2 c_1 \alpha_Y$, where $\mu_{5,Y}^i$ is the initial chiral asymmetry. 
This mechanism works when the chirality flip interaction by electron Yukawa interaction is ineffective 
when the instability grows. This gives the lower bound of the initial chiral asymmetry for this mechanism 
to work as $\mu_{5,Y}^i /T_i \gtrsim 10^{-3\sim 4}$. 
As a result, hyper magnetic field properties around the electroweak symmetry breaking and today are given by~\cite{Kamada:2018tcs,Schober:2017cdw} 
\begin{align}
B_Y(T_{\rm EW}) &\simeq 0.82 {\rm GeV}^2 \ c_1^{-1/3}  \left(\frac{\mu_{5,Y}^i/T_i}{10^{-2}}\right)^{1/3} \left(\frac{T_{\rm EW}}{10^2 {\rm GeV}}\right)^{7/3}, \label{BT}\\
\lambda_Y(T_{\rm EW}) &\simeq 9.8 \times 10^6 {\rm GeV}^{-1} \  c_1^{-1/3} \left(\frac{\mu_{5,Y}^i/T_i}{10^{-2}}\right)^{1/3} \left(\frac{T_{\rm EW}}{10^2 {\rm GeV}}\right)^{-5/3},  \label{lamT} \\
B_0 \simeq & 9.9 \times 10^{-16} {\rm G} \ c_1^{-1/3} \left(\frac{\mu_{5,Y}^i/T_i}{10^{-2}}\right)^{1/3} , \\ \quad 
\lambda_0  \simeq&  6.9  \times 10^{-3} {\rm pc} \ c_1^{-1/3}  \left(\frac{\mu_{5,Y}^i/T_i}{10^{-2}}\right)^{1/3} . 
\end{align}
Thus with relatively large $\mu_{Y,5}^i/T_i$ and not too large $c_i$ (or relatively large right-handed electron asymmetry), 
the relatively large magnetic fields can be generated in this mechanism (but at relatively small scales). 

\section{GUT Baryogenesis as the source of large initial chiral asymmetry}

Then the question is the origin of such a large initial chiral asymmetry. 
One possibility is SU(5) GUT (Higgs) boson in the ${\bm 5}$ representation decay into $e_R u_R$ and ${\bar Q}_L^1 {\bar Q}_L^1$, 
which can produce initial chiral asymmetry and initial right-handed electron asymmetry. 
In the usual GUT baryogenesis through the thermally produced ${\bm 5}$ Higgs decay, 
we could not have large chiral asymmetry and right-handed electron asymmetry 
to avoid the monopole problem and stronger couplings to the 2nd and 3rd generations of fermions. 
However, if we suppose a mechanism to produce the ${\bm 5}$  scalar nonthermally such as instant preheating,
and the ${\bm 5}$  scalar is not related to the electroweak symmetry breaking but couples to the first generation tighter, 
we can have large initial chiral asymmetry and right-handed electron asymmetry with $c_1 = 553/481$~\cite{Kamada:2018tcs}. 

Note that the hyper magnetic fields are maximally helical and produced before the electroweak symmetry breaking, 
as clarified in Refs.~\cite{Fujita:2016igl,Kamada:2016eeb,Kamada:2016cnb}, baryon asymmetry is (re)produced at electroweak 
symmetry breaking through the chiral anomaly in the SM. 
Then it turns out that the magnetic fields suggested by the blazar observation predict baryon overproduction if the 
magnetic fields are maximally helical. 
Thus, the magnetogenesis mechanism that predicts the generation of maximally helical hyper magnetic fields 
before the electroweak symmetry breaking cannot be responsible for the blazar observation. 
In the present mechanism, the present baryon-to-entropy ratio is predicted 
in terms of the initial chiral asymmetry and right-handed electrons as~\cite{Kamada:2018tcs}
\begin{align}
\eta_B^0 \simeq &4.0 \times 10^{-5} c_1^{-1}  \left(\frac{\mu_{5,Y}^i/T_i}{10^{-2}}\right)  f(\theta_W, T), 
\end{align}
where $f(\theta_W, T) ={\cal O}(10^{-3} \sim 1)$ is the numerical coefficient that depends on the detail of the electroweak symmetry 
breaking. 
However, by an appropriate choice of parameters, it can be responsible for the present baryon asymmetry of the Universe. 
In other words, although SU(5) GUT baryogenesis has been thought not to work since it does not produce $B$-$L$ asymmetry 
and the electroweak sphalerons wash out the asymmetry generated there, 
it is revived through the chiral plasma instability. 
This is because the electroweak sphalerons cannot wash out the asymmetry carried temporarily by the hyper magnetic fields.

\section{Summary}

In this presentation, we discussed the importance of the chiral plasma instability in the early Universe cosmology. 
The recent observations of TeV blazars motivate us to examine the possibility that 
relatively large magnetic fields existed in the early Universe. 
The chiral plasma instability is one of the possible mechanisms that can work in the early Universe, which has plenty of interesting phenomenology. 
In the present study, it turns out that due to the baryon overproduction problem, 
unfortunately, it cannot be responsible for the intergalactic magnetic fields suggested by the blazar observations. 
However, it can be the source of the baryon asymmetry of the Universe today. 
In particular, through this mechanism, SU(5) GUT baryogenesis might be revived as its indirect origin, 
which otherwise was thought not to work due to the washout by the electroweak sphalerons.

\section*{Acknowledgements}
The work of KK was supported by IBS under the project code, IBS-R018-D1.

\end{document}